\documentclass[]{aa}
\usepackage{graphicx}
\usepackage{txfonts}
\usepackage {natbib}
\usepackage[dvips]{epsfig}
\newcommand{\msolyr}{\ifmmode{{\rm M}_{\odot}~{\rm 
yr}^{-1}}\else{{M$_{\odot}$~yr}$^{-1}$}\fi}
\newcommand{\msun}{\ifmmode{{\rm M}_\odot}
\else{M$_{\odot}$} \fi}
\newcommand{\lsun}{\ifmmode{{\rm L}_{\odot}}
\else{L$_{\odot}$} \fi}
\newcommand{\rsun}{\ifmmode{{\rm R}_{\odot}}
\else{R$_{\odot}$} \fi}
\newcommand{\zsun}{\ifmmode{{\rm Z}_{\odot}}
\else{Z$_{\odot}$} \fi}
\newcommand{\teff}{\ifmmode{{\rm T}_{\rm eff}}
\else{T$_{\rm eff}$} \fi}
\newcommand{\mdot}{\ifmmode{\dot{\rm M}} 
\else{$\dot{\rm M}$}\fi} 
\begin{document}

\title{Low and intermediate mass star yields.\\
 II: The evolution of nitrogen abundances}
\author{Marta Gavil\'{a}n\inst{1},  
 Mercedes Moll\'{a} \inst{2} and James F. Buell\inst{3}}

\offprints{Marta Gavil\'{a}n}

\institute{Departamento de F\'{\i}sica Te\'{o}rica, Universidad Aut\'onoma 
de Madrid, 28049 Cantoblanco, Spain\\
\email{marta.gavilan@uam.es}\\
\and Departamento de Investigaci\'{o}n B\'{a}sica, 
C.I.E.M.A.T., Avda. Complutense 22, 28040 Madrid, Spain \\ 
\email{mercedes.molla@ciemat.es}\\
\and Department of Mathematics and Physics, Alfred State College, 
Alfred, NY 14802, USA \\ 
\email{BuellJF@alfredstate.edu}\\
}

\date{Received ; accepted }

\titlerunning{Low and intermediate mass star yields}
\authorrunning{Gavil\'{a}n,  Moll\'{a} \& Buell }

\abstract
{}
{We analyze the impact on the Galactic nitrogen abundances of using a
new set of low and intermediate mass star yields. These yields have a
significant yield of primary nitrogen from intermediate mass stars.}
{We use these yields as an input to a Galactic Chemical Evolution
model and study the nitrogen abundances in the halo and in the disc,
and compare them with models obtained using other yield sets and with
a large amount of observational data.}
{ We find that, using these new yields, our model adequately reproduce
the observed trends. In particular, these yields solve the historical
problem of the evolution of nitrogen, giving the right level of
relative abundance N/O by the production of a primary component in
intermediate mass stars.  Moreover, using different evolutionary rates
in each radial region of the Galaxy, we may explain the observed N
dispersion.}
{}

\maketitle

\section{Introduction}

Most elements are created in the interiors of stars by nucleosynthesis
processes \citep[see][for a review]{wall97}, starting with hydrogen
and progressing toward heavy elements. These processes are called {\sl
primary production}.  Some elements, however, can be formed from
nuclei heavier than hydrogen originally present in the star.  They are
called {\sl secondary}. This is the case of nitrogen, that can be
created during the CNO cycle using seeds of original carbon and/or
oxygen.  From a theoretical point of view, it has been considered that
massive stars produce secondary nitrogen \citep{pei87}, while low and
intermediate mass (LIM) stars have mechanisms, like the third
dredge-up and the Hot Bottom Burning processes, to produce both,
primary and secondary nitrogen \citep{edm78,all79}.  The third
dredge-up event is a consequence of the thermal pulses in the star,
and transport C and He to the outer layers. The Hot Bottom Burning
occurs when the CNO cycle takes place at the base of the convective
envelope.

Observationally, there are several open questions about the primary or
secondary character of nitrogen that up to now remain unsolved. When N
and O data are represented as log(N/O) {\sl vs} log(O/H), including
the galactic stars, H{\sc ii} regions for the Milky Way Galaxy (MWG),
external galaxies \citep{gar95,gar99,vze98,izo99} and the high
redshift data \citep[][and references therein]{pet02,pro02,cen03}, a
clear positive slope appears for abundances larger than $\rm
12+log(O/H)=7.8-8$ dex which indicates a secondary behavior, but the
plot shows a flat slope for low metallicities that can only be
explained with a primary component of nitrogen. Taking into account
that this flat slope occurs for low abundances, the first idea
proposed, shared by some authors \citep{pag79,dia86,dah95}, is that
observations would be reproduced if the nitrogen ejected by massive
stars would be primary, while intermediate mass stars might have both
primary and secondary components.

Thus, some authors have tried to look for mechanisms that explain how
massive stars could produce primary nitrogen. This is the
case of \citet{mey02} that have recently proposed rotation as a
possible source of primary nitrogen, since low metallicity stars show
a bigger rotation than high metallicity ones. \cite{chi03-1} have used
these yields in their chemical evolution models, concluding that they
are only a lower limit for the primary nitrogen production since the
Hot Bottom Burning is not considered in their calculation.  In fact,
\cite{chi05} find that an extra-- production of N in low metallicity
massive stars by a large factor, between 40 and 200 along the mass
range, is necessary to explain the data of very metal-poor halo stars
since these yields do not produce a sufficient amount of primary
N. Moreover, if the production of primary nitrogen would proceed from
massive stars, the left side of the (N/O) {\sl vs} (O/H) plot should
not show any scatter.  Although some authors claim to observe
\citep{izo99,pil03} this no-scatter, recent observations from low
metallicity objects \citep{pet02,pro02,cen03,isr04,spi05} do show a
clear dispersion.

\citet{ser83} already claimed that a secondary production by
intermediate mass stars must exist and suggested that the zero slope
may be explained by two factors: 1) a delay in the ejection of N to
the ISM due to the different mean-lifetimes of stars and 2) the gas
infall effects. The advantage of taking a delay into account is that
the great data scatter can be explained by considering different
evolutionary states for each galaxy and, therefore, this possibility
has been supported by a large number of authors
\citep{vila93,pil92,pil93,vze98b,hen00}. These last ones also include
gas flows --infall or/and outflow--, and low efficiency for the star
formation rate (the equivalent mechanism to produce a delay) in the
low evolved regions, in order to reproduce the flat slope in the (N/O)
vs (O/H) plot.  They conclude that the secondary production of
nitrogen should dominate in high metallicity environments while the
primary one should act at low metallicities.  

Some new yields for LIM stars have been given in \citet[][hereinafter
Paper I]{gav05}, where they were adequately evaluated and calibrated
by using them in a Galaxy chemical evolution model. It was shown that
the results about C and O abundances adequately reproduce the Galactic
and Solar Neighborhood data. The purpose of this work is to analyze
the impact of these stellar yields on the nitrogen abundances. In
particular, we check, using the same prescriptions of Paper I, if the
contribution to N given by these yields for LIM stars is sufficient to
justify the amount of primary nitrogen the observations point out.

We describe the yields in Section 2, analyzing in particular the
primary and secondary components of the nitrogen production. In
section 3 we describe briefly the chemical evolution model. Section 4
is devoted to the results, and the conclusions are presented in
section 5.

\section{Low and intermediate mass yields: 
The secondary and primary components of nitrogen}
\label{prim-t-sec}

The aim of this work is the study of the nitrogen behavior, using the
same set of yields as in Paper I, that we call BU yields.  For
comparison purposes we also take the LIM stars yields from
\cite{hoe97} and \cite{mar01} that we call VG and MA, respectively.
The complete table of yields BU was already given in Paper I for five
metallicities: -0.2, -0.1, 0.0, +0.1 and +0.2, expressed as $\rm
log(Z/Z_{\odot})$, where solar abundances are taken from \cite{gre98}
\footnote{The use of these solar abundances implies that
$Z_{\odot}=0.02$. Recently, \cite{asp05} have obtained lower
abundances which lead to a value $Z_{\odot}=0.012$. However, these new
determinations are still questioned by some authors
\citep{bah05,dra05,ant05} because they do not fit the
helioseismological constraints.}.

\begin{figure}
\resizebox{\hsize}{!}{\includegraphics[]{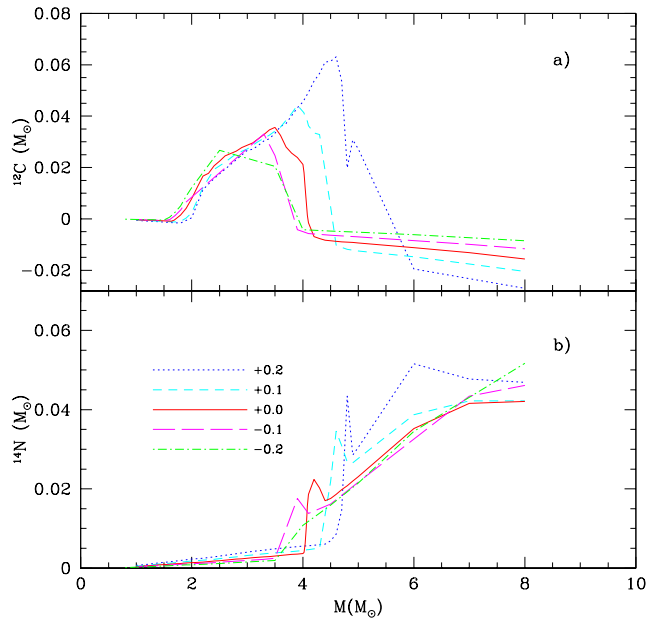}}
\caption{Total yields (BU) of: a) $^{12}C$ and b) $^{14}N$ 
produced by LIM stars for different metallicities following label 
on panel b), expressed as $\rm log{(Z/Z_{\odot})}$. }
\label{yields}
\end{figure}

We summarize the behavior of the carbon and nitrogen yields for LIM
stars, as shown in Fig.~\ref{yields}.  In panel a) we see that
$^{12}C$ yield is extremely small for stars with mass lower than 2
$\rm M_{\odot}$, since they do not experience third dredge up events.
However, stars begin to suffer these kinds of events for smaller
masses at lower metallicity.  In other words, in the low mass range,
the metallicity and the $^{12}C$ yield are anti-correlated.  When the
stars have enough mass to undergo Hot Bottom Burning (HBB), the
$^{12}C$ yield drops abruptly because of the conversion of carbon into
nitrogen.  The $^{14}N$ yield presents a local maximum in the mass
range from 3.5 to 5 M$_{\odot}$, depending on the metallicity, then
decreases before beginning to increase again as a function of stellar
mass.  The largest amount of nitrogen is produced by stars of
intermediate mass because HBB and the 2$^{\rm nd}$ dredge-up occur
only in stars with $M>3.5-5 M_{\odot}$.  As the HBB increases the
luminosity and the mass-loss rate, stars that suffer this process have
shorter TP-AGB lifetimes. The local maximum occurs in the transition
between stars with HBB and those without. The increase at higher
masses is due to the shortened time between third dredge-up
events. The yields at the lowest masses are due to the 1$^{st}$
dredge-up.
\begin{figure}
\resizebox{\hsize}{!}{\includegraphics[angle=-90]{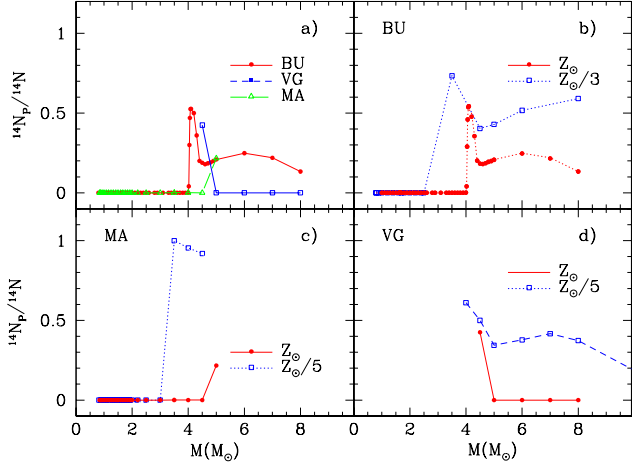}}
\caption{The ratio of the yield of primary $^{14}{\rm N}$ to the total
yield of $^{14}{\rm N}$ as derived from different author sets: BU, VG
and MA.  Panel a) a comparison of the solar metallicity yields for the
three sets as labeled.  Models for two different compositions, as
labeled, are represented in panels b), c) and d) for BU, MA and VG,
respectively.}
\label{prim_total}
\end{figure}


The most important difference among the used yields resides in the
contribution of primary and secondary components of nitrogen by LIM
stars. In Fig.~\ref{prim_total} we represent the fraction of primary
$^{14}N$ for the three used sets, as labeled, as a function of mass
(M $\leq$ 8 $M_{\odot}$, except for MA for which M $\leq 5$
$M_{\odot}$). In panel a) we show the results for solar abundances.
All of them show a similar behavior with a maximum for masses around
3.5-4 $M_{\odot}$.


We must clear some points about the components of N.  The only
difference between primary and secondary nitrogen is the origin of the
carbon atom producing it. Although the idea is conceptually clear, it
is not so simple to separately compute both components. Thus, although
BU and MA give the two components separately for each model, VG do
not. These authors, however, show their yields in each phase of
stellar evolution.  If we consider that all the nitrogen created in
the AGB phase is primary, about $ \sim 90$ \% of the N ejected by
LIM stars will be primary. This is sometimes assumed when
these yields are used. This hypothesis, that we call {\it AGB
technique} leads to a primary N component excessively large and is not
totally adequate.

Let us return to the definition: secondary N proceeds from the burning
of original $^{12}C$. If a fraction of the original carbon is burned
in the pre-AGB phase, it produces secondary N.  Sometimes, this gives
a negative $^{12}C$ yield. But, not all the initial carbon is consumed
before the AGB phase.  If we take, as an example, a star of
4$M_{\odot}$ of solar abundance, that is with $X(^{12}C)=0.28\times
10^{-2}$, it has an initial $^{12}C$ abundance $4 \msun X(^{12}C) =
1.12 \times 10^{-2} \msun$ The pre-AGB phase carbon yield is
$yC12_{pre} = 0.300\times 10^{-4}$, so the mass of this element
present in the star before the AGB begins is:

\begin {equation}
M(^{12}C)=yC12_{pre}M_{ini}+M_{end} X(^{12}C)
\end{equation}

where $M_{end}$ is the mass of the star at the end of this first
phase: 3.95 $\msun$. Therefore, there is a mass $\sim 1.118\times
10^{-2} M_{\odot}$ of $^{12}C$, from which $M_{end}X(^{12}C) \sim
1.102\times 10^{-2}M_{\odot}$ corresponds to original carbon. This
implies that a quantity of the original carbon is still available to
form nitrogen in the following phases. Thus, a fraction of the total
nitrogen produced in the AGB phase (given by the addition of the two
values given by VG in their tables denoted AGB yields and final AGB
yields) may be secondary.  In order to calculate this component from
the total AGB yields we use the fraction, called $r$, between the
secondary to the total nitrogen yield, $r=^{14}N_{S}/^{14}N$.  Taking
into account that the secondary N proceeds from the existing carbon
used as a seed, we assume that $r$ is equal to the ratio between the
old carbon and the new plus old carbon:

\begin{equation}
r= \frac{^{14}N_{S}}{^{14}N} =
\frac{^{12}C_{old}}{^{12}C_{old}+^{12}C_{new}}
\end{equation}  

This method (hereinafter called $r$ method), may only be applied to
stars which suffer the HBB and produce primary $^{14}N$, that is,
those for which the core mass before the HBB is larger than
$M_{HBB}=0.8 M_{\odot}$, usually stars with $M> 3.5-4
M_{\odot}$. Otherwise, the nitrogen yield is all secondary.  The
results of VG shown in Fig.~\ref{prim_total} proceed from this
calculation.
 
\begin{figure}
\resizebox{\hsize}{!}{\includegraphics{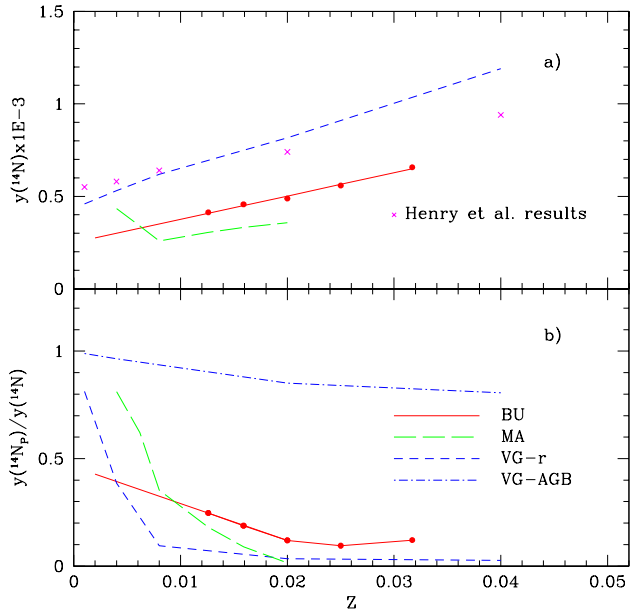}}
\caption{a) Dependence of the integrated yield of $^{14}N$ produced by
LIM stars on metallicity for the three yield sets, marked with
different symbols as in Fig.~\ref{prim_total}. The crosses are the
results obtained by \cite{hen00} for the same VG yields using a
Salpeter IMF. b) The ratio between the primary to the total nitrogen
yield with the same symbols than in panel a). The two possible
techniques to compute the primary component of VG yields are
represented as short-dashed line (method $r$) and dot-short-dashed
line (method $AGB$).}
\label{yields_integrados}
\end{figure}

We have then computed the integrated yields of $^{14}N$ produced by
LIM stars that we present in Fig.~\ref{yields_integrados} as a
function of metallicity Z.  In panel a), we represent the BU results
as solid circles to which we have performed a least-squares fit shown
by the solid (red) line. This integrated yield for $^{14}N$,
equivalent to the yield produced by a single stellar population, is
located between the two other sets in this panel, with a similar
dependence on Z that VG \footnote{The integrated yields for VG are
slightly different than those obtained by \cite{hen00} due to the
Initial Mass Function (IMF) used by us, from \cite{fer92}} but with
lower absolute values.

More significant however, is the dependence on metallicity of the
ratio of primary to total integrated yields, $^{14}N_{P}/^{14}N$,
shown in panel b). This ratio increases for decreasing metallicity for
all sets, as expected, although for VG the integrated yield is quite
different if we consider AGB technique than $r$ method.

This metallicity effect can be easily explained: low metallicity stars
have smaller radii and take longer to reach super-winds, so they have
more time to experience more third dredge-up events than solar
metallicity stars. As a consequence, they have more fresh $^{12}{\rm
C}$ in their envelopes and they can make more primary nitrogen by the
HBB process. On the other hand, due to the lower amount of original
carbon, they produce less secondary nitrogen. For VG yields the ratio
is almost constant at a value of 15\% when the $r$ technique,
represented by the short-dashed (blue) line, is used, for
metallicities greater then 0.01, although it also increases for
metallicities lower than this value.  While it is $ \sim 90$ \% when
the $AGB$ technique, represented by a dot-short-dashed (blue) line, is
used, showing a smaller variation with Z. It is interesting that the
integrated yield for solar abundance in the BU case \footnote{For $Z<
0.0126$ we use the same yields set, that is, this one from
$Z=0.0126$. It is always possible to extrapolate the trend obtained
for the other Z sets, that we show with the straight line joining the
points corresponding to $Z=0.0126 $ and $Z=0.0159$} is around 20\%,
very similar to the value computed by \citet{all79} two decades ago on
the basis of the observations available at that time.

All these considerations indicate that the primary nitrogen appears at
a different time scale in the ISM depending on the value of Z. The first
primary N will be ejected when stars of 8 (5 for MA) $M_{\odot}$ die,
while Z is still low.  
 
\section{The chemical evolution model and its calibration}
\subsection{Description}

The model used in this work is the Multiphase Chemical Evolution Model
described in \citet{fer92,fer94}, in the version presented in
\citet{mol05} and in Paper I. For LIM stars, we use the same yields BU
than in these two last papers, and for comparison purposes those from
MA and VG. For massive stars we have chosen \cite{por98} and
\citet[][hereinafter PCB and WW, respectively]{woo95}. We have run
different models computed with different combination of yields: BU +
WW, VG + WW and MA + PCB, which we distinguish as BU, VG and MA,
respectively.

The nitrogen study is usually done by comparing its behavior relative
to iron and oxygen, so it is very important to have a careful
calibration for these two elements. Oxygen calibration was done in
Paper I. The SNIa are the main manufacturer of iron. The yields for
type Ia supernova (SNIa) explosions are taken from \cite{iwa99} and
\cite{bra86}.  The evolution of this element in the model is
quasi-independent of the normal stars yields. However, since iron is
mainly produced by SNIa, even if its yield is very well known, its
abundance is very dependent on the method to compute the rate of these
explosions. For this purpose we analyzed the results obtained with
different possibilities in order to eliminate, if possible,
uncertainties in the iron abundance evolution. This point is
relatively important because the Age-metallicity relation and the
G-dwarf metallicity distribution are usually used as calibration
methods for chemical evolutions models. In our case, furthermore, we
compared our results with observed stellar nitrogen abundances, most
of which are given as [N/Fe], and so, we checked that the Iron
evolution is adequately reproduced by our models before this
comparison can be made
 
We used three methods to compute the SNIa rates as given by the
following authors: the classical one \citep{mat86,fer93}, the one
given by \cite{tor89}, and other, more recent, described in
\cite{ruiz00}, hereinafter named MAT, TOR and RL, respectively. The
first authors estimate the SN rates by using only the Initial Mass
Function. The method, well described in depth in both cited works, is
summarized as follows: a proportion of the stellar masses in a given
range [$M_{min}$-- M$_{max}$] will be in binary systems and a fraction
of them will develop type Ia supernova.  Based on this idea, a mass
function for the secondary stars is computed from the original
one. Finally the SNIa rate depends on the number of secondary stars
that died in each time step, which implies that the time scale for the
iron appears in the ISM is controlled by the mean lifetimes of these
secondary stars.

\begin{figure}
\resizebox{\hsize}{!}{\includegraphics[angle=-90]{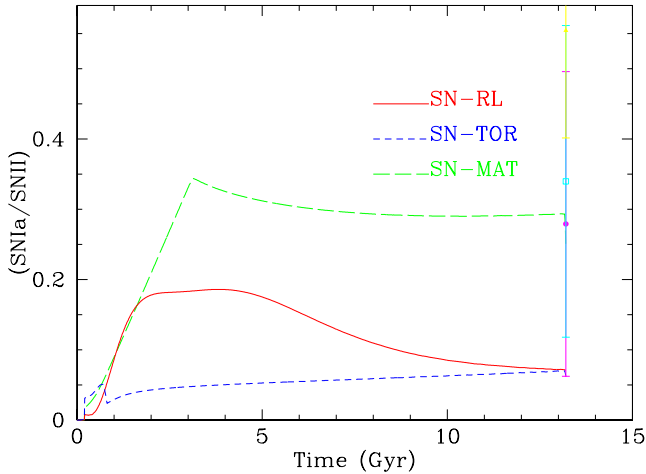}}
\caption{Evolution of SNIa/SNII rates for MAT, TOR and RL techniques as
labeled in the figure. The observed values for the Galaxy, given by
\citet{capp99,capp04,mann05}, are shown, with the error bars, by the
magenta full dot, the cyan open square and the black triangle,
respectively.}
\label{tasas}
\end{figure}

Actually, this time scale does not depend only on the secondary mean
lifetimes, since there are other processes that also participate in
the conversion of a binary system into a SNIa explosion. It is
necessary to take into account the effects of the distances between
both stellar components, the orbital velocities and other parameters
to finally obtain the time taken for the system to explode since the
moment of its formation.  \cite{tor89} performed these calculations
for several combinations of possible candidates of binary system or
SNIa scenarios (Double Degenerate, Single Degenerate, etc...),
providing the supernova rate as a function of time normalized for a
binary system of 1 $\rm M_{\sun}$.  All the physical processes and
assumptions are included in their calculations, so we only need to
include the selected functions in our code and multiply them by the
number of binary systems, avoiding the need of computing the secondary
and primary initial mass function as defined in the previously
described method. A similar technique has also been performed more
recently by \cite{ruiz00}. A numerical table has been provided to us
by Ruiz-Lapuente ( private communication) with the time evolution of
the supernova rates for a single stellar population, computed under
updated assumptions about different scenarios and probabilities of
occurrence.  We have computed the supernova rates using the three
methods, thus producing three models MAT, TOR and RL. These different
techniques affect mostly the iron abundances, the other elemental
abundances being equal for all of them. Therefore, we will compare the
three type of SN rate calculations by using only BU yields in the
analysis of the Iron abundance evolution, as well as in the
calibration of the model (next section). We will compare the three set
of LIM stars yields when N be studied, using only the RL technique and
only for the comparison of the relative abundance [N/Fe] we will show
the nine possible combinations of models.

\begin{figure}
\resizebox{\hsize}{!}{\includegraphics{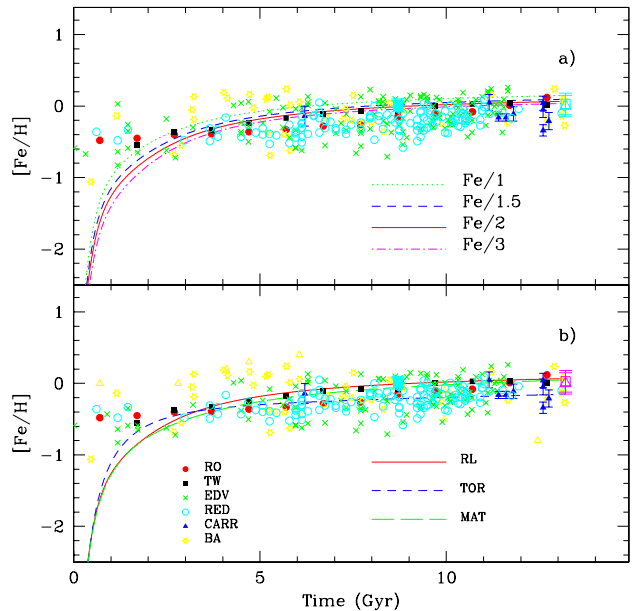}}
\caption{Age-Metallicity relation: a) for BU and RL model dividing WW
iron yield by 1, 1.5, 2 and 2.5. b) using MAT, TOR and RL SNIa rates
with BU yields. Data are from authors of Table~\ref{authors} as
labeled.}
\label{AMR}
\end{figure}

The main disparity among the three techniques described above resides
in the different evolution of the SN rate in time. As we see in
Fig.~\ref{tasas}, MAT is the technique that presents highest values of
SNIa/SNII at any time, reaching the maximum at 2.5 Gyr. RL has a
maximum between 2 and 5 Gyr, with values approximately 1/2 or 1/3 of
those given by MAT. Nevertheless it still is within the error bar
given by observations \citep{capp99,capp04,mann05}. Note that this
value has been reduced for the most recent determinations compared
with the oldest ones.  TOR model is the only one with low values.
Even if it presents a maximum before the first Gyr, this will not be
seen in the results because its value is very small. From the first
Gyr, SNIa/SNII has positive slope and it almost reaches the observed
value at the present time.  \footnotesize
\begin{flushleft}
\begin{table*}
\begin{tabular}{lccccccc}
\hline
\noalign{\smallskip}
Reference  &  Fe & C   & N  & O    & R     & Age \\    
\noalign{\smallskip}
\hline
\noalign{\smallskip}
\cite{ake04}(AKE)       &  X      &  X  & --- &   X  & --- & --- \\
\cite{bar88}	     	&  X      & --- & --- &   X  & --- & --- \\
\cite{bar89}	     	&  X      & --- & --- &   X  & --- & --- \\
\cite{barry88}(BA)   	&  X      & --- & --- &  --  & --- &  X  \\
\cite{boe99}	     	&  X      & --- & --- &   X  & --- & --- \\
\cite{car87}(CARB)   	&  X      & X   & X   & ---  & --- & --- \\
\cite{car98}(CARR)    	&  X      & --- & --- & ---  & --- &  X  \\
\cite{car00-2}	     	&  X      & X   & X   &   X  & --- & --- \\
\cite{cav97}	     	&  X      & --- & --- &   X  & --- & --- \\
\cite{che00}	     	&  X      & --- & --- &   X  & --- &  X  \\
\cite{cle81}	     	&  X      & X   & X   &   X  & --- & --- \\
\cite{daf04}(DAF)    	& ---     & X   & X   &   X  & X   & --- \\
\cite{dep02}	     	&  X      & X   & X   &   X  & --- & --- \\
\cite{ecu04}	     	&  X      & --- & X   &  --- & --- & --- \\
\cite{edv93}(EDV)    	&  X      & --- & X   &   X  & X   & X   \\
\cite{gus99}	     	& ---     & X   & --- &  --- & X   & X   \\
\cite{fri90}	     	&  X      & X   & --- &  --- & --- & --- \\
\cite{gra00}	     	&  X      & X   & X   & X    & --- & --- \\
\cite{gum98} 	     	& ---     & X   & X   & X    & X   & --- \\
\cite{isr98,isr01}   	& X       & --- & --- & X    & --- & --- \\
\cite{isr04}(ISR)    	& X       & --- & X   & X    & --- & --- \\
\cite{lai85}	     	& X       &  X  & X   & ---  & --- & --- \\
\cite{mel01},	     	& X       &  X  & --- & ---  & --- & --- \\
\cite{mel02}	     	&         &     &     &      &     &     \\
\cite{mis00}	     	& X       & --- & --- &  X   & --- & --- \\
\cite{nis02,nis02b}     & X       & --- & --- &  X   & --- & --- \\
\cite{red03}(RED)       & X       & X   &  X  & ---  & --- &  X  \\
\cite{roc00,roc00b}(RO) & X       & --- & --- & ---  & --- &  X  \\
\cite{rol00},	        & X       & X   &  X  & X    & X   & --- \\
\cite{sma97},	        &         &     &     &      &     &     \\
\cite{sma01}	        &         &     &     &      &     &     \\
\cite{shi02}	        & X       & X   &  X  & ---  & --- & --- \\
\cite{smi01}	        & X       & --- & --- & X    & --- & --- \\
\cite{spi05}(SPI)       & X       & X   &  X  & X    & --- & --- \\
\cite{tom84},	        & X       & X   &  X  & X    & --- & --- \\
\cite{tom86,tom95},     &         &     &     &      &     &     \\ 
\cite{twa80}(TW)        & X       & --- & --- & ---  & --- & X   \\ 
\cite{wes00}            & X       & X   &  X  & X    & --- & --- \\
\hline
\noalign{\smallskip}
\end{tabular}
\caption{References for CNO stellar abundances used for the comparison
with model results.}
\label{authors}
\end{table*}
\end{flushleft}
\normalsize

\subsection{Calibration of the model: Iron evolution in the Solar Vicinity}

\begin{figure}
\resizebox{\hsize}{!}{\includegraphics{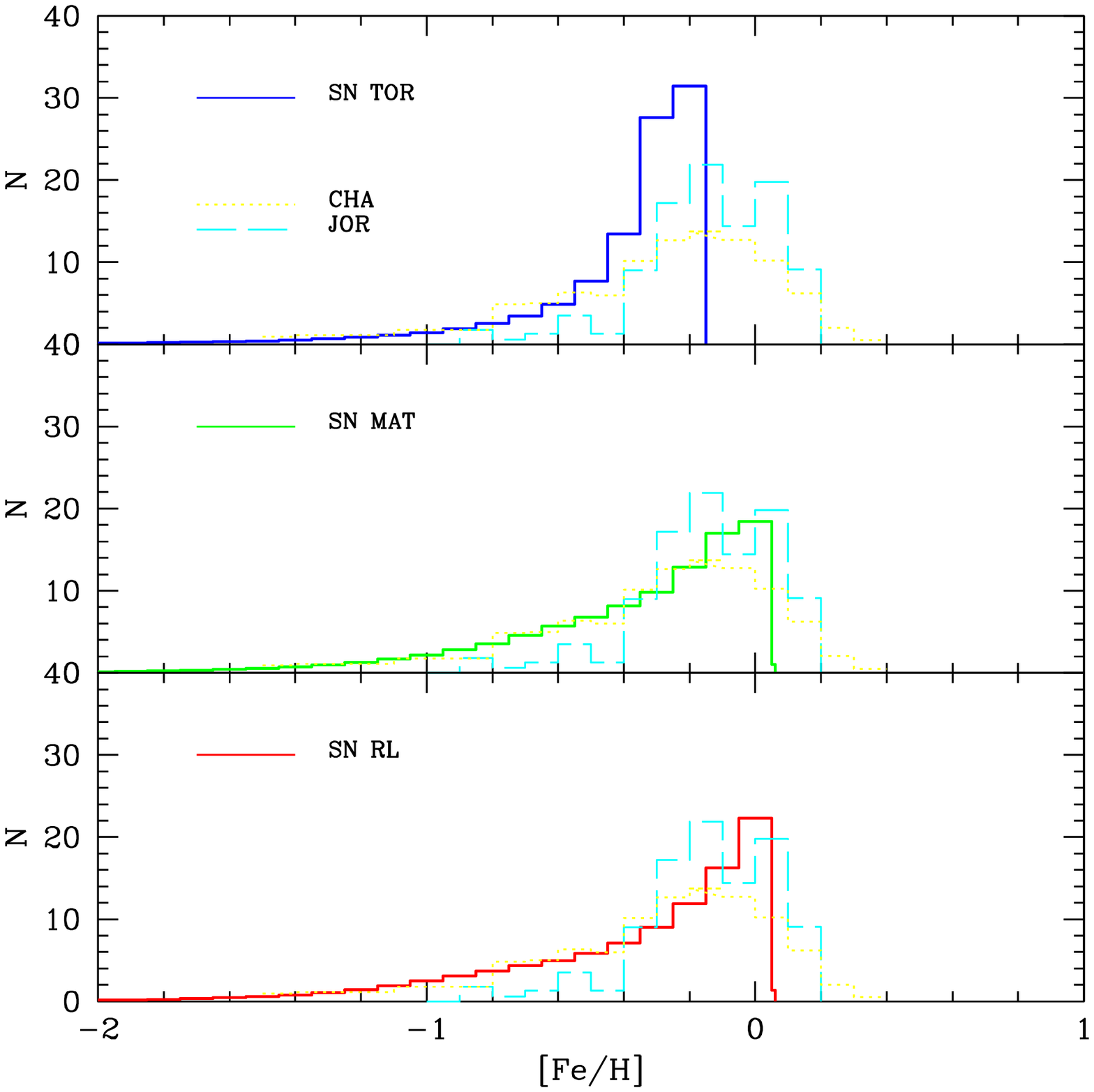}}
\caption{The G-Dwarf distribution in the Solar Vicinity. Data are from
\cite{cha00} (dashed line) and \cite{jor01}, (long-dashed line)}
\label{G_dwarf}
\end{figure}

The results for iron abundance obtained with these three methods are
shown in Fig.~\ref{AMR}, the Age-Metallicity Relation (AMR) and in
Fig.~\ref{G_dwarf}, the G-Dwarf distribution, for the Solar Vicinity.
For this comparison we have shown only BU yields, keeping in mind that
the set of yields will have only small effects on this relation.
Nevertheless, this model uses WW yields for massive stars and these
authors claimed that this could produce too much iron, and advised in
\cite{tim95} to divide the iron ejections at least by two. In order to
calculate how much that WW iron excess will be, we have calculated
four different models for BU yields and RL technique, where massive
stars iron production is divided by 1, 1.5, 2 and 3. Results are
represented in panel a) where it is clearly shown that a factor of 2
is a good compromise that we will use in panel b).  In this last panel
we present the Age-Metallicity relation for the three SNIa cases,
where all of them are in reasonable agreement with data, given their
large dispersion.  Although there are very little differences between
models, it can be seen that the iron appears later and takes a little
more time to reach high values when MAT and RL techniques are used,
than for the TOR SNIa method, but all of them reach the solar
abundance.

Regarding the G-dwarf distribution, represented in Fig.~\ref{G_dwarf},
differences appear mainly between TOR and the others because it
provides a narrower distribution than the others.  The three models
are able to reproduce the low metallicity tail without showing any
G-dwarf problem.  

\begin{figure}
\resizebox{\hsize}{!}{\includegraphics{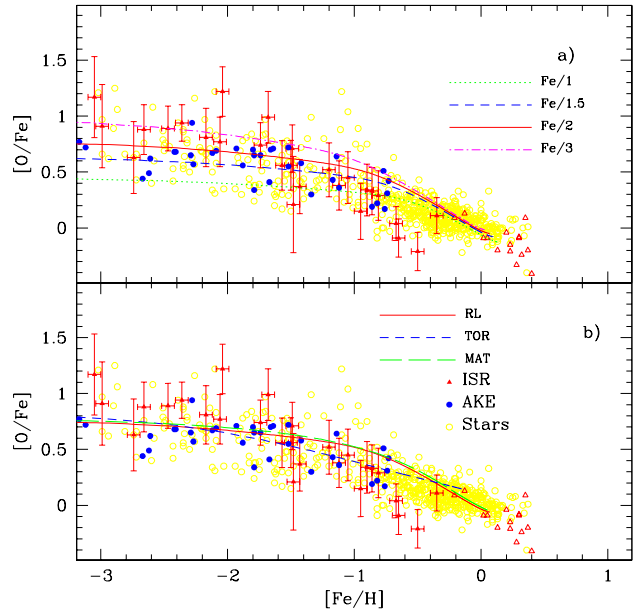}}
\caption{The relation [O/Fe] {\sl vs} [Fe/H]: a) The model results for
BU + RL varying WW iron ejection; b) model BU using MAT, TOR and RL
SNIa techniques and BU yields. Open (black) dots are stellar data
from authors listed in Table ~\ref{authors}, from which we represented
the most recent from \cite{isr04} and \cite{ake04} as (red) triangles,
and full (blue) dots. }
\label{feo}
\end{figure}

In Fig.~\ref{feo} we show the relation between iron and oxygen. As
before, in panel a) the BU + RL model is presented varying the massive
stars iron ejection.  In this case the differences are clearer than in
the AMR case.  We chose the model Fe/2 that we will use for the rest
of the paper. In panel b) we plot the model results using BU yields
with the three SNIa techniques. As in the previous case, the LIM stars
yields do not change the results because oxygen is ejected by massive
stars and iron is mainly produced by SNIa events. The three models
have very similar behavior.

\section{Results analysis: The nitrogen abundances}

We devote this section to analyzing the results obtained with our
models for nitrogen abundances and comparing them with the available
observational data. We divide these results in four parts: a) the
evolution of Nitrogen over time for the Solar Vicinity (assumed
located at a galactocentric distance of 8 kpc), b) the radial
distributions of elements in the disc, c) the relation of log(N/O)
with the oxygen evolution and d) the relation of [N/Fe] with iron.

\subsection{Time evolution of nitrogen}
\begin{figure}
\resizebox{\hsize}{!}{\includegraphics{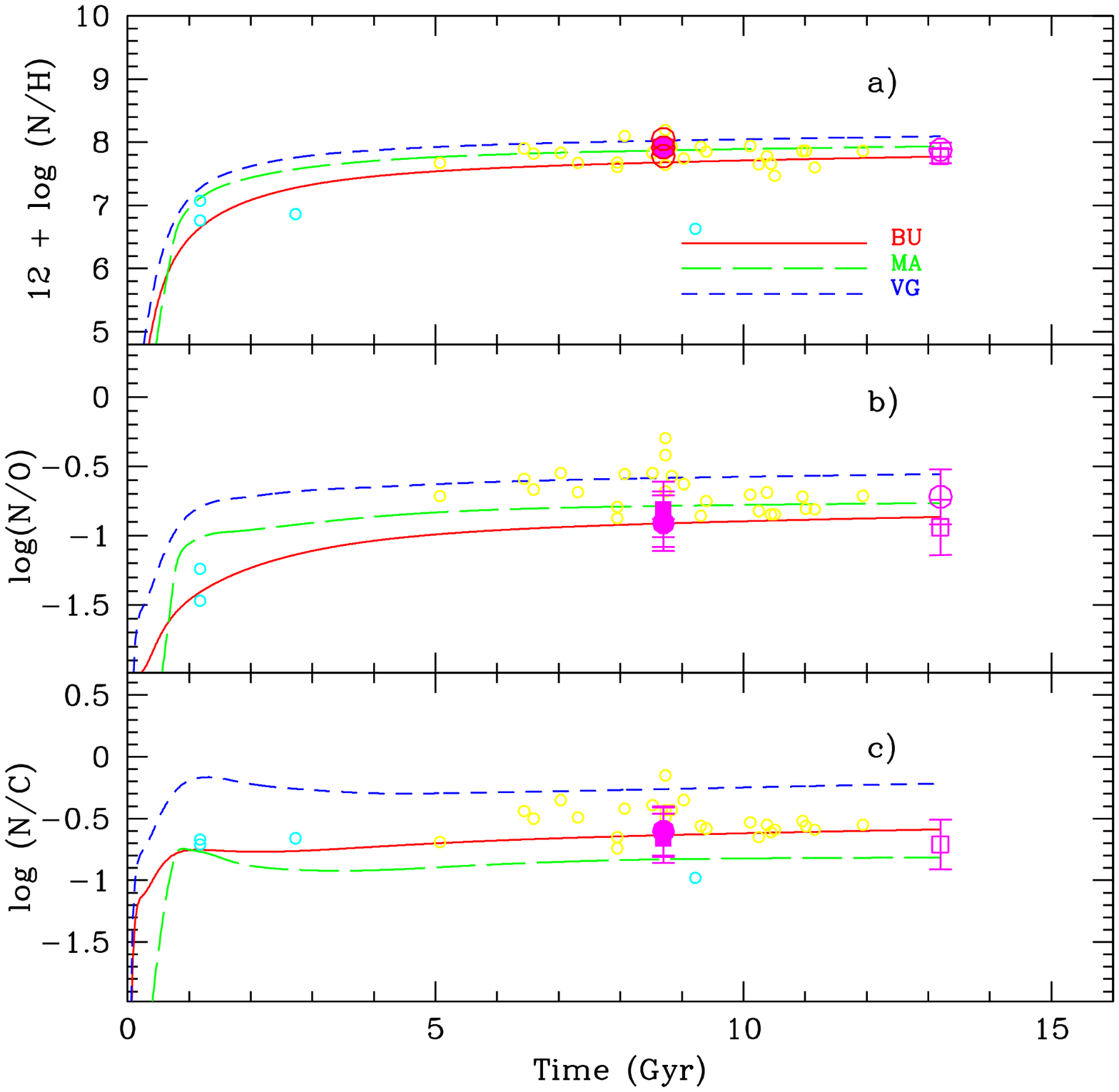}}
\caption{Time evolution of elemental abundances in Solar Vicinity for
 a) nitrogen, as $12+log(N/H)$; b) nitrogen over oxygen as log(N/O)
 and c) nitrogen over carbon as log(N/C).  Solar abundances are the
 filled symbols from \cite{gre98} -- circle--, \cite{hol01} --square--
 and \cite{asp05}, --cross--, by assuming an age of 4.5 Gyr for the
 sun.  Empty symbols at 13.2 Gyr are the interstellar medium
 abundances given by \cite{mey97,mey98}, --circle-- and \cite{pei99}
 --square--. Small open dots are stellar abundances obtained by
 authors from Table~\ref{authors}, being those located around the
 Solar Vicinity ($7.5 < R < 9.5$ kpc). Lines meaning is given in panel
 a). }
\label{abunt}
\end{figure}

Fig.~\ref{abunt} shows the evolution of nitrogen with the solar
abundance values --large filled symbols-- taken from \cite{gre98} --
circle--, \cite{hol01} --square-- and \cite{asp05}, --cross--, by
assuming an age of 4.5 Gyr for the sun. For the interstellar medium
abundances at 13.2 Gyr, large empty symbols, we use the abundances
given by \cite{mey97,mey98}, --circle--, and \cite{pei99}
--square--. The small open circles are abundances for objects with
given stellar ages, at a radial galactocentric distance between 7.5
and 9.5 kpc.

Models BU, VG and MA are represented by the (red) solid, the (blue)
short-dashed and the (green) long-dashed lines, respectively.  Both
models VG, following the two possible techniques to calculate the
proportion of primary nitrogen, techniques $r$ and $AGB$ described in
Section~ \ref{prim-t-sec}, yield results indistinguishable for times
larger than 1 Gyr, so we represent only the results for the first one.
In panel a) MA and VG models give a greater value than BU since at the
lowest metallicity their nitrogen yield are higher than the
corresponding one from BU (see Fig.\ref{prim_total}). Then, once an
abundance higher than $\sim$0.004 is reached, it continues increasing
smoothly until the present time, reproducing both, solar and ISM,
abundances. The shape shown by the three models are similar and all of
them reproduce both the solar value and the ISM value.

The same kind of information can also be extracted from the relative
abundances represented in panel b). In panel b), we show the
time evolution of log(N/O).  Since there is a good agreement in
fitting the abundance of oxygen for all models, \citep[see][]{gav05},
the differences in this plot must be only due to the nitrogen
production. The disagreement between the different models is important
for times shorter than 1.5 Gyr, when intermediate mass and massive
stars are the main contributors and the distinct primary/secondary ratios
effects are evident there. Model MA has a strong increase in the first
Gyr due the primary component and then it flattens up. Model BU has a
ratio of primary nitrogen larger than MA for all Z except for the
lowest one, what produces a smoother evolution, and, finally, N
remains below MA. The resulting final N/O ratios are similar in both
models and in agreement with observations. The behavior mostly primary
of model VG, when all AGB nitrogen is considered as such, implies a
very strong increase of the abundance at the earliest times. After
that, both methods give a smooth slope, reaching an absolute value
around -0.5 dex higher than the observed one.

\begin{figure}
\resizebox{\hsize}{!}{\includegraphics[angle=0]{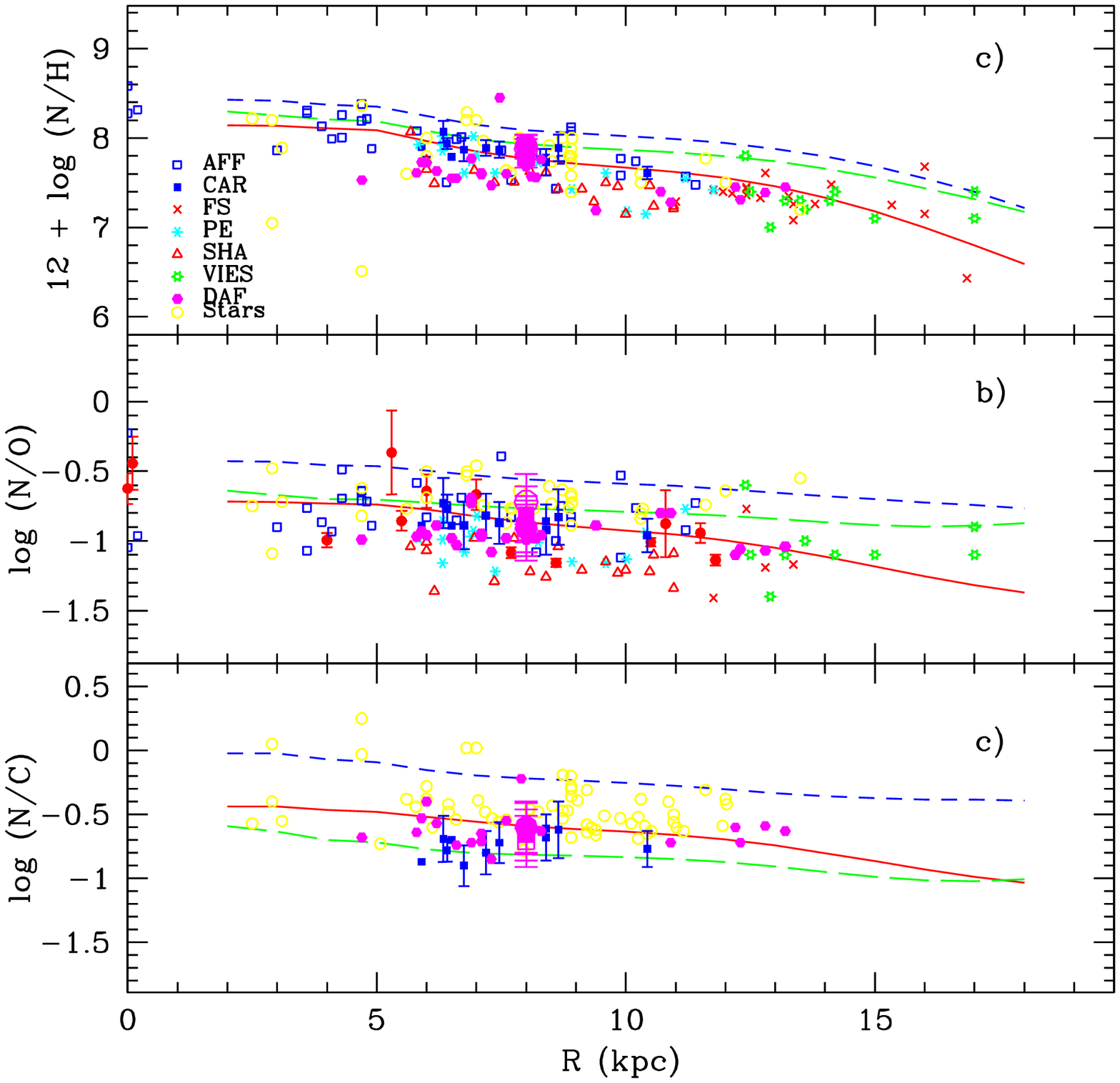}}
\caption{Radial distributions of nitrogen abundances, a) as 12 + log
(N/H), b) as log(N/O) and c) as log(N/C) for the same models than the
previous figure.  Stellar data --open dots-- are taken from authors
given in Table~\ref{authors}, including the most recent ones from
\cite{daf04} marked separately.  Galactic H{\sc ii} regions abundances
are also included, taken from \cite{pei79} --PE--, \cite{sha83}
--SHA--, \cite{fic91} --FS--, \cite{vil96} --VIES--,
\cite{aff97}--AFF-, and \cite{est99,est05,car05} --CAR-- as marked in
the figure. The solar and interstellar abundances have the same symbol
than in Fig.~\ref{abunt}.}
\label{gran}
\end{figure}

The good behavior of BU yields is also evident in panel c) where
log(N/C) is shown. MA model presents a maximum at the first Gyr, the
decrease is due to the higher amount of carbon ejected in that model
(see Paper I); so the absolute value at the present time is only
marginally reached. The shape for VG model is similar than the BU one
but with a nitrogen excess. All models seem to fit the solar and ISM
data but the model BU is the best one in reproducing the stellar data,
and, more importantly, it is the best one at fitting all of the data
at the same time.

\subsection {Nitrogen abundance in the Galactic disc}
Now, we will explore the radial distributions of nitrogen over the
galactic disc, shown in Fig.~\ref{gran}.  Data correspond to H{\sc ii}
regions from references as labeled in the figure and to stars from
references in Table~\ref{authors}.

The radial distribution is more or less reproduced within the errors
by all models.  Actually, the shape of the radial distribution is well
fitted in all cases, independently of the absolute values, since this
results as an effect of the ratio infall/SFR along the galactocentric
radius, produced by the scenario of our MWG model, and therefore is
rather independent of the yields used.  However, the slope of the
radial distribution at the two ends of the disc, in the center and in
the outer regions, is a matter of discussion. Thus, \cite{vil96} claim
that the gradients are not as steep in these regions as in the rest of
the galactic disc. The same occurs in the inner disc where the most
recent data from \cite{sma01} show that the distribution flattens. Our
models have been tuned to fit these two sets of data, and due to that,
the resulting overall gradient is smaller than that obtained by other
authors.  As can be seen in Fig.~\ref{gran} MA and VG models produce a
gradient flatter than observed. 

In Fig.~\ref{gran}b) and c), the radial distributions log(N/O) and
log(N/C) are plotted, as they are considered important for the study
of different yields. In panel b) the radial distribution of log(N/O)
showed by data presents a clear slope, although there is some data
that shows a flatter distribution in the outer regions. A steep radial
distribution for N/O is expected because oxygen is produced by massive
stars.  If nitrogen was ejected by massive stars, its secondary
character would cause it to enter the ISM after the oxygen. Instead,
if it would be ejected by intermediate stars, the time needed for
their evolution would be larger. Thus, in both cases, the nitrogen
appears in the ISM after the oxygen does.

Once again, the large dispersion of the data prevents a clear
selection of {\sl the best model}, however the BU model seems most
adequate to reproduce the H{\sc ii} regions data from \cite{vil96} and
\citet{fic91}. The MA model shows a radial gradient flatter than
indicated by observations, with higher absolute values compared to the
mean values of data, mostly in the outer disc, and VG models, as
before, show higher values.  The same arguments are also valid when
panel c) is analyzed. In this case a slight negative gradient is shown
for log(N/C). The small amount of data and the great dispersion
prevent the selection of any model as better than the others, although
it seems clear that MA remains below most of them as corresponds to
the large production of C, and that VG lies in the upper side of the
data.  It is apparent that BU model shows a better behavior compared
to the data.  Once again we stress the importance of using adequate
yields to reproduce the whole set of data at the same time. Yield BU
seems to be in the adequate range of production of N, C and O, since
the model appears in the zone occupied by of data in the three panels.

It is necessary to remember that open dots represent stellar
abundances. We have tried to select only those corresponding to young
stars, but we do not know the age of the complete set of stars with
available data.  In this case we have preferred to use the available
abundances; thus, it is possible that some data does not correspond to
young enough stars.

\subsection {Nitrogen {\sl vs} Iron}

In this case, as the iron evolution may have also an influence over
the model results, we have presented the relation between nitrogen
and iron, Fig.~\ref{nfe}, with a different panel, a), b) and c), for
each set of yield, MA, BU and VG, respectively.  The three possible
methods to compute the SNIa, RL, TOR, and MAT are shown with solid
(red), short-dashed (blue) and long-dashed (green) lines,
respectively, in each panel.  The first thing we observe is that the
effects of the different SNIa techniques are almost indistinguishable.
Therefore, the main features of each model at those metallicities are
due to yields. In other words, we may analyze the behavior of the
nitrogen corresponding to each yield set disregarding the accuracy in
the SNIa calculations.

\begin{figure*}
\resizebox{\hsize}{!}{\includegraphics{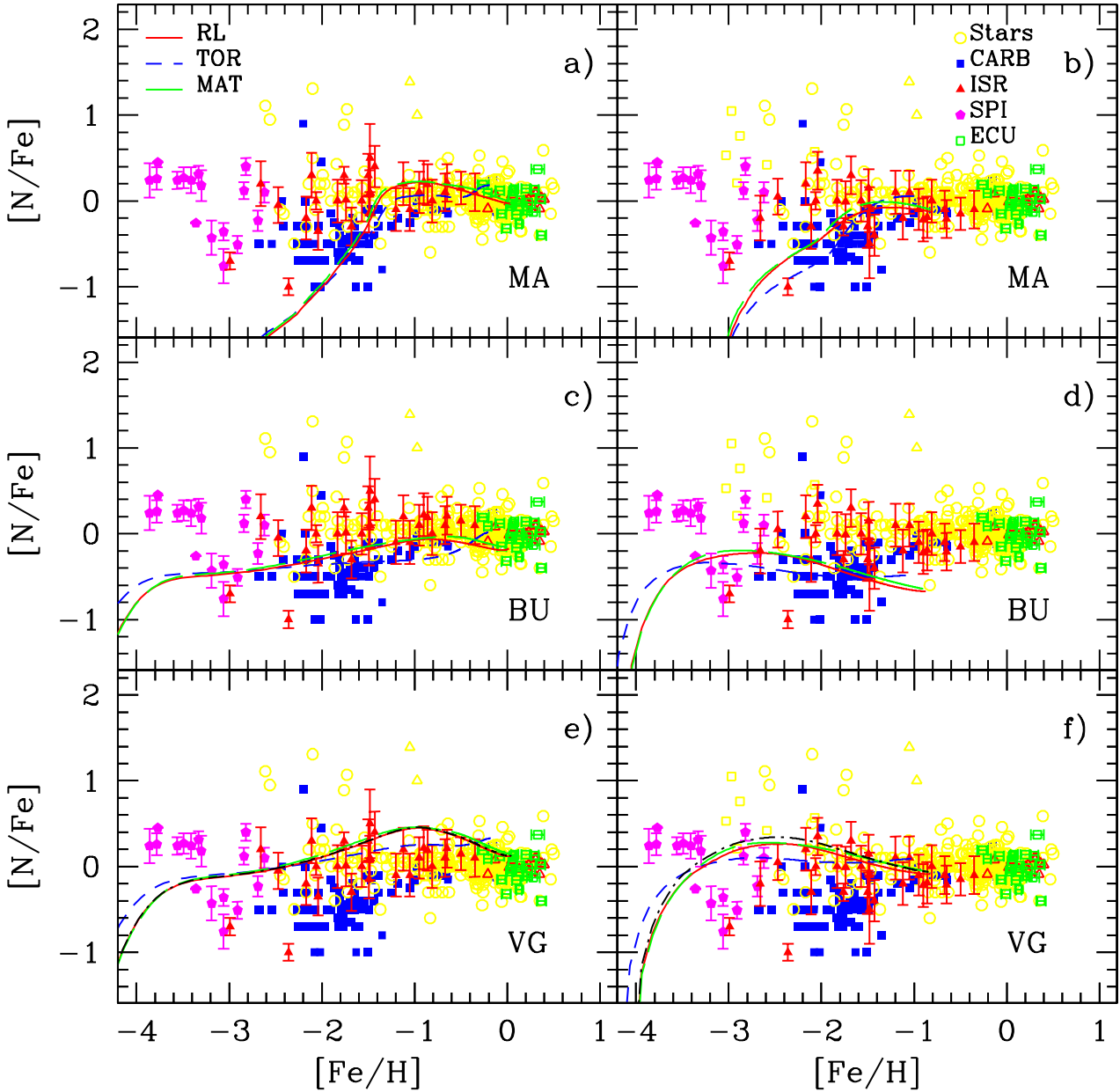}}
\caption{The relative abundances of [N/Fe] {\sl vs} the iron abundance
[Fe/H] for the solar region.  The panels on the left show the disc
evolution and the panels on the right show the halo evolution. In each
one of them, the evolution obtained with three different methods to
compute the SN-Ia rates are shown as labeled in panel a).  Data are
taken from the different references listed in Table~\ref{authors}
marked separately the most recent ones from \cite{isr04,ecu04,spi05}
in panel d).  }
\label{nfe}
\end{figure*}
\label{nfe}

All results agree in the sense that the first nitrogen to be ejected
is secondary, as due to the massive stars, so the initial slope is
positive and large, --although this behavior is not shown in the
figure because it occurs when oxygen abundances is lower that 5 dex--
but they differ in when the slope begins to change. When the N ejected
by LIM stars appear, there is a strong increase due to the change from
a secondary to a primary behavior.  In MA yields LIM stars eject less
primary nitrogen, and later, since it is ejected as secondary for
stars up to 5 $M_{\odot}$. Then the main contributors to primary N are
the stars with masses between $2 M_{\odot}$ and $3 M_{\odot}$. For
this reason the slope does not change until it reaches $\rm [Fe/H] =
-1.5$, the moment in which these stars begin to die.  When BU yields
are used, the trend changes earlier in the evolution due to the
contribution of the primary nitrogen ejected by stars in the range
4--8 $M_{\odot}$. As their life is so brief, the ejection occurs at
$\rm [Fe/H] = -4$. From then, the slope is close to zero: the
signature of primary nitrogen.  The case of VG shows a behavior
similar to BU.

We would like to remark that the data dispersion is so great that all
the models lie in the data area, regardless of their big
discrepancies, although the region of the metal-rich objects ($[Fe/H]
> -1.5$) is particularly well fitted in panel b) by Model BU. It is
necessary to use the very low metallicity data to clarify which model
works better. In fact, the most recent observations from
\cite{isr04,spi05} show a slope flatter than before, (even with a
negative slope) which is a behavior more consistent with our model BU
than with the model obtained with MA yields. This last model might be
considered acceptable when the available low metallicity data were
only those from \cite{car87}, but when using the new determinations of
N abundances for this kind of objects, the conclusion is that BU
better reproduces the generic trend of data.  It is also necessary to
take into account that most of the metal-poor objects do not belong to
the disc but to the halo. In this way, we represent the halo model
results for the zone that infall over the disc at galactocentric
distance equal to 8 Kpc, at the right panels b), d) and f) of
Fig.\ref{nfe}.  We see that the trend shown by the recent observations
from \cite{isr04,spi05} is more compatible with BU and VG than MA.

\subsection {Nitrogen {\sl vs} oxygen}

\begin{figure}
\resizebox{\hsize}{!}{\includegraphics[angle=-90]{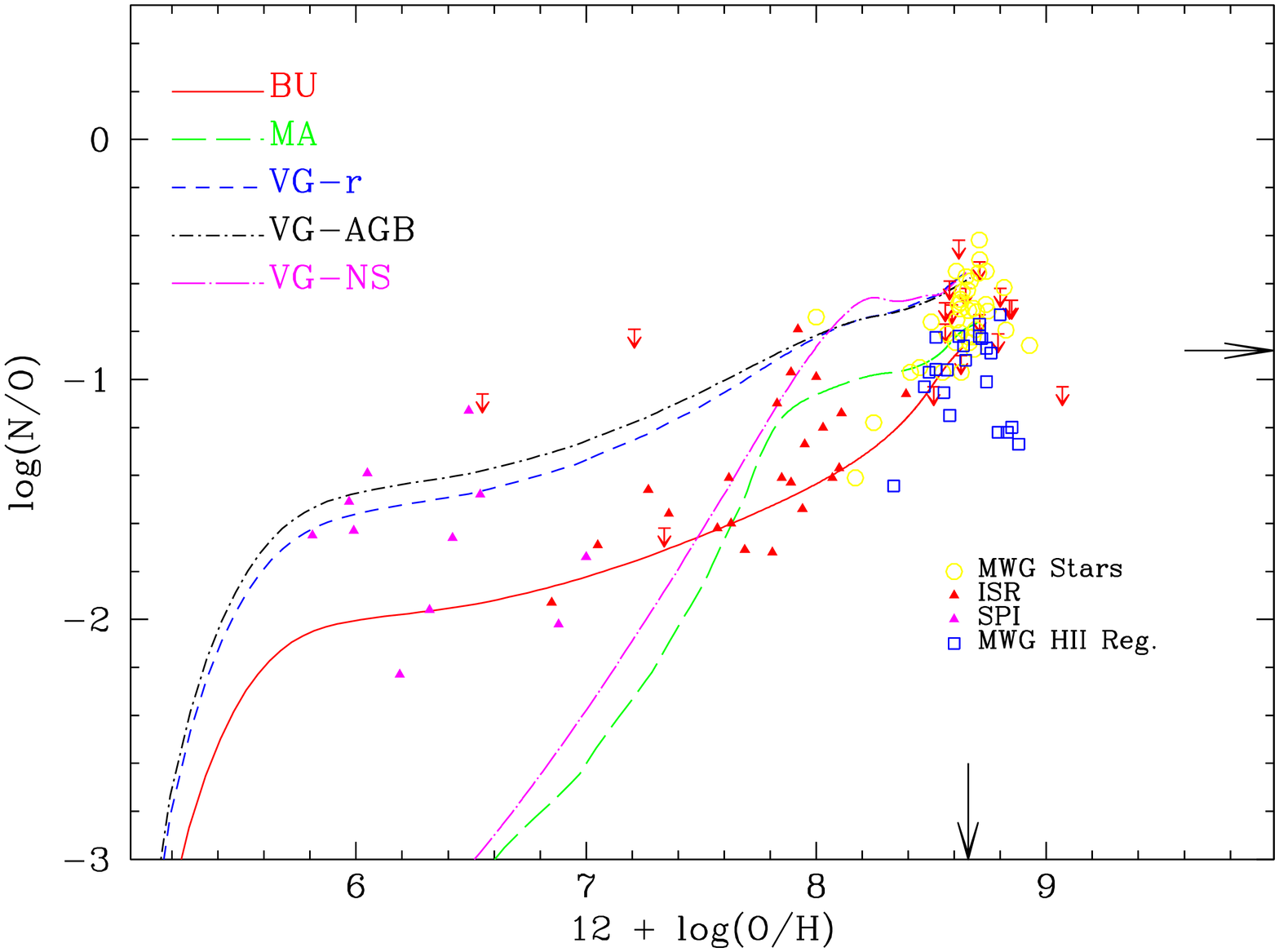}}
\caption{The relative abundances of log (N/O) {\sl vs} the Oxygen
abundance as $12+log(O/H)$ for the Solar Neighborhood (lines as shown
in the plot).  Stellar abundances are the open dots from
Table~\ref{authors}, from which we mark separately those from
\cite{isr04} and \cite{spi05} as (red and magenta) full triangles.
Galactic H{\sc ii} regions data in the Solar Region ($7.5 < R(kpc) < 9.5$
are from \cite{pei79}, \cite{sha83},
\cite{fic91}, \cite{vil96}, \cite{tsa03} and \cite{car05}. The meaning of
symbols is given in the figure. The solar position is marked with
arrows. }
\label{no}
\end{figure}

Finally, we show in Fig.~\ref{no} the classical and well known graphic
of the relative abundance of N {\sl vs} O as $\log{(N/O)}$ {\sl vs}
$12 +\log{(O/H)}$.  We show the final results for the Solar
Neighborhood of the computed models. Model BU reproduces well the
expected behavior of N when the whole data is taken into account.  Not
only the level of N is so adequate but also the shape is smoother than
the one shown by the other two models.  The observed trend at low
metallicity can be well reproduced with BU yields because they have
the appropriate primary to secondary ratio, and the adequate
integrated nitrogen yields. MA yields have also a primary nitrogen
component, but the integrated nitrogen yield has a metallicity
dependence in the opposite way as BU for the lowest Z, so the trend
shown by data can not be well reproduced.  Both VG models have the
right shape and are almost the same for $12 +\log{(O/H)} \ge 8$.  The
problem is that the integrated yields are high. It would be necessary
to change the input parameters as the infall rate or the efficiencies
to form stars in order to fit the solar abundances. In that case,
probably, other data will not be reproduced. Only the BU model shows
simultaneously the good shape and adequate absolute abundances.  For
comparison purposes, we have also shown the resulting model using VG
yields but assuming that the Nitrogen is completely secondary.

If the flat behavior were caused by massive stars, the data
dispersion would be very small. A problem arises when the metal-poor
objects \citep{isr04,spi05} are included in the figure, as can be seen
in Fig.~\ref{no}. Some values follow the described trend over the flat
line, but there exist some lower abundances which are around
$log(N/O)\sim -2$. This behavior is not compatible with a primary
component proceeding only from massive stars.

\begin{figure}
\resizebox{\hsize }{!}{\includegraphics[angle=-90]{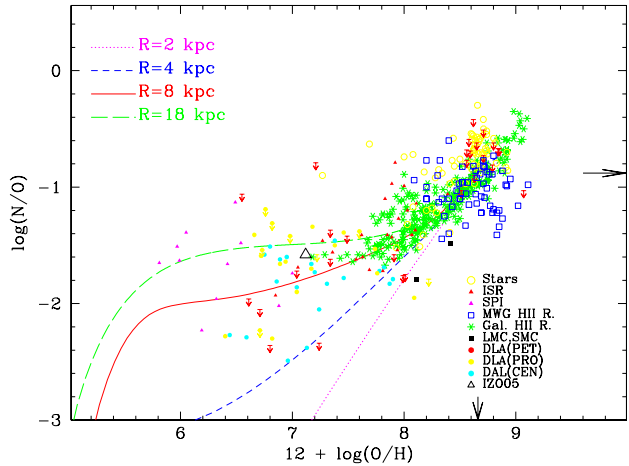}}
\caption{The relative abundances of log (N/O) {\sl vs} the Oxygen
abundance as $12+log(O/H)$ for the halo --upper panel-- and for the
disc --lower panel-- for four different galactocentric
regions: two inner regions located at 2 and 4 kpc, dotted and short
dashed lines, a solar region, solid line, and an outer region, long
dashed line, respectively. Data refer to Galactic stars are from the same
authors as cited in Table~\ref{authors}. The meaning of symbols is
given in the figure. The solar position is marked with arrows. }
\label{no_bis}
\end{figure}

We show in Fig.\ref{no_bis} the evolution given by BU model for four
different radial regions of the Galaxy: two inner ($\sim$ 2 and 4 kpc)
more evolved regions --(magenta) dotted and (blue) short-dashed
lines--, the solar vicinity ($\sim$ 8 kpc) as in the previous figure,
the (red) solid line, and an outer one ($\sim$ 18k pc), --the (green)
long-dashed line--, where the evolution takes place slowly. In panel
a) we show the results for the halo zones and in panel b) for the disk
regions. In both panels we have included the stellar data for the MWG.

The numbers on the graph indicate the evolutionary time, in million
years, that corresponds to that point of the line. This is necessary
because the $12+log(O/H)$ value is not the same for each radius at the
same value of time.

The halo regions have similar evolutions independent of their distance
from the center of the Galaxy. All of them reproduce well the recent
data from \cite{isr04,spi05} obtained for $5.5 < 12+log(O/H) < 8$, and
the disk regions fit the stellar data obtained for $12+log(O/H) > 8$
of the disk.  Their evolutionary tracks, however, are very different,
as corresponds to their distinct input parameters (infall rates,
initial gas masses, efficiencies to form stars...) which are
translated into very different star formation histories. Thus, the
dispersion of the MWG data can be well explained on the basis of a
primary production of nitrogen from LIM stars, higher for the lowest
metallicities, and with different star formation efficiencies in the
different regions.

\begin{figure}
\resizebox{\hsize }{!}{\includegraphics[]{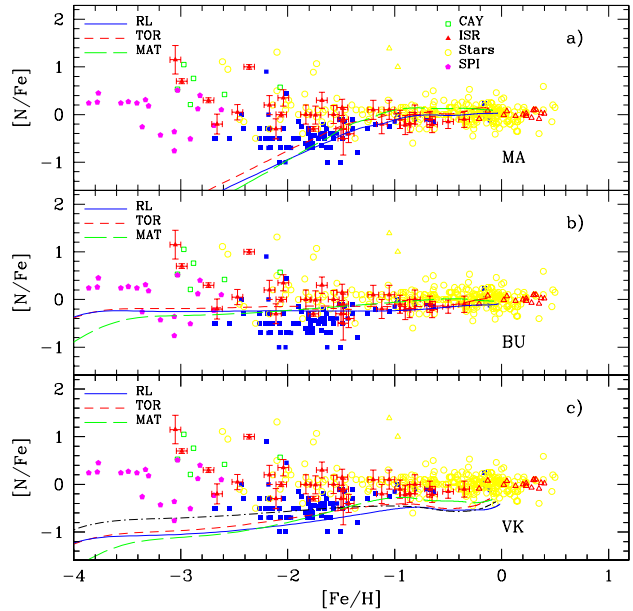}}
\caption{The relative abundances of log (N/O) {\sl vs} the Oxygen
abundance as $12+log(O/H)$ for the same disk regions than previous
figure compared with data referring to Galactic and extragalactic HII
regions and DLA objects data. The meaning of symbols is given and the
solar position is marked with arrows. } \label{no_hii}
\end{figure}

We represent the same results for the disk regions in Fig.\ref{no_hii}
compared with data referring to Galactic HII regions, taken from the
same authors than those of Fig.~\ref{no}, but without limiting the
possible galactocentric distance.  Other galaxy data
\citep{gar95,gar99,vze98} and \cite{izo99} are also shown.  The large
open triangle is the recent estimate obtained from \cite{izo05} for
the lowest metallicity known galaxy. We have also added the DLA
objects data from \cite{pet02,pro02,cen03} as solid points.  We want
to remark that the disk regions evolve in good agreement with all of
them, showing the inner regions a steeper evolution while the outer
one shows a very flat evolution with a high and constant value
$log(N/O)\sim -1.2$ dex, similar to the behavior of dwarf galaxies.
These results suggest that the observed dispersion in this kind of
plot, when other galaxies data (such as dwarf or DLA galaxies) are
included, might be reproduced if different star formation histories
have occurred in different galaxies.  This argument has already been
invoked by other authors, in particular by \cite{hen00} and
\cite{pra03}. It was even demonstrated by \cite{pil03}, who analyzed
data for different radial regions in spiral galaxies and showed the
changes of the evolutionary track in the plane N/O {\sl vs} O/H for
each one of them. It is evident that this kind of behavior may be
represented by our models and that the new yields may reproduce better
the whole set of data. In fact, these yields have already been used in
a grid of chemical evolution models for a large number of theoretical
galaxies \cite{mol05}. A discussion about the resulting N/O abundances
and its possible dispersion for different objects is done in
\cite{mol05b}.

This figure and the behavior of (C/O) vs O/H, shown in Paper I, are
the main clues to consider the present yields the most adequate to
represent the evolution of galaxies. The production of carbon by LIM
stars is sufficient to obtain an increase in C/O without the need to
invoke mass loss by massive star winds, and the N/O behavior may be
well reproduced with different star formation efficiencies due to the
adequate level of the primary component produced by LIM stars and to
the right dependence of this component with Z.

\section{Conclusions}
Our conclusion can be summarized as follows:

\begin{enumerate}
\item The primary component of nitrogen, necessary to explain the
trend of N/O with O/H, may be mostly produced by LIM stars, and
adequately fits all the data, including the observed dispersion. In
this way the integrated yield produced by LIM stars must be directly
proportional to Z, while the ratio $N_{P}/N_{tot}$ must increase for Z
decreasing. A primary component, larger for lowest metallicities, has
an important effect on explaining low abundance range data.
\item The dependence of the N yield on stellar mass would have a
maximum around 5-6 $M_{\odot}$, while the primary component shows
other around 3.5-5 $M_{\odot}$. This constraints the time where these
contributions have important effects on the evolution.
\item The high dispersion on N/O data for low and high metallicity
galactic regions may be explained with these yields as we have
demonstrated with the radial regions of the disc models which have
different star formation efficiencies. Our findings go in the same
address than \cite{hen00} and \cite{pra03} using VG yields, but we
claim that it is easier to reproduce the whole data set when BU yields
are used. Our model for MWG with small efficiencies to form stars is
consistent with data. 
\item These results seem suggest that models with differences in the
star formation histories for different types of galaxies, such as
those calculated in \cite{mol05}, might produce final abundances
with high dispersion, in agreement with the observed one when dwarfs
galaxies or DLA galaxies are included in a plot N/O-O/H, such as we
will show in \cite{mol05b}.
\item As \cite{chi03-2}, we also support that the halo and the disc
have different evolutions.  The set of stellar data [C/Fe], [N/Fe] and
[N/C] may be divided into two trends. The first one is well reproduced
by our disc models, while the second one is well fitted by our halo
results.
\item Due to the primary N component of BU yields, and since the
intermediate stars have short lifetimes, it is possible to produce
high [N/Fe] abundances even at low metallicities which are in perfect
agreement with the recent halo stars data obtained by
\cite{isr04,cay04,spi05}.
\item We claim that the fit of the whole set of data with only one
model is not an easy task.  We may reproduce the observed trend with
BU yields combined with yields from WW. In summary, our model BU
reproduce reasonably well the whole CNO data set.
\end{enumerate}

\begin{acknowledgements} 
This work has been partially supported by the Spanish PNAYA project
AYA2004--8260-C03-03. We acknowledge Pilar Ruiz-Lapuente for her
personal contribution in the SNIa rates data.  We also thanks Jose
Manuel V\'{\i}lchez for his valuable suggestions and the referees,
Leonid S. Pilyugin and Angeles I. D\'{\i}az their comments that have
improved this paper.
\end{acknowledgements}
 
\bibliographystyle{aa}  
\bibliography{bibliografia}
\end{document}